\newcommand{\manuscript}{0}

\ifnum\manuscript=1
\documentclass[12pt,preprint,flushrt]{aastex}
\else
\documentclass{emulateapj}
\fi

\usepackage{epsfig}
\usepackage{bm}

\newcommand{\etal}{et al.\,}
\newcommand{\usr}{\ $\mu$sr}

\newcommand{\x}      {\ensuremath{\bm{x}}}
\newcommand{\bfu}    {\ensuremath{\bm{u}}}
\newcommand{\bfv}    {\ensuremath{\bm{v}}}
\newcommand{\bfcdot} {\ensuremath{\bm{\cdot}}}
\newcommand{\bfDelta}{\ensuremath{\bm{\Delta}}}
\newcommand{\etarj}  {\ensuremath{\eta_{RJ}}}
\newcommand{\dif}    {\ensuremath{\mathrm{d}}}
\newcommand{\ra}[3]  {\makebox[1.5em][r]{#1}\makebox[1.5em][r]{#2} \makebox[2em][r]{#3}}

\shorttitle{CMB Observations with MINT}
\shortauthors{J. W. Fowler \etal}

\begin{document}

\title{CMB Observations with a Compact Heterogeneous 150 GHz
Interferometer in Chile}
\author{
J.~W.~Fowler,
W.~B.~Doriese,\altaffilmark{1,6}
T.~A.~Marriage, 
H.~T.~Tran,\altaffilmark{2,7}
A.~M.~Aboobaker, 
C.~Dumont,\altaffilmark{3}
M.~Halpern,\altaffilmark{4}
Z.~D.~Kermish,\altaffilmark{2}
Y.-S.~Loh,\altaffilmark{5}
L.~A.~Page,
S.~T.~Staggs,
D.~H.~Wesley
} 
\affil{Department of Physics, Jadwin Hall, Princeton University,
P. O. Box 708, Princeton, NJ 08544.} 
\altaffiltext{1}{Present address: NIST Quantum Electrical Metrology Division, 325
Broadway Mailcode 817.03, Boulder, CO 80305-3328.}
\altaffiltext{2}{Present address: Department of Physics, University of California,
Berkeley, CA 94720.}
\altaffiltext{3}{Present address: School of Medicine, University of
Pennsylvania, Philadelphia, PA 19104.}
\altaffiltext{4}{Department of Physics and Astronomy, University of
British Columbia, Vancouver, BC, Canada  V6T 1Z4.}
\altaffiltext{5}{Present address: Center for Astrophysics and Space Astronomy,
389 UCB, University of Colorado, Boulder, CO 80309.}
\altaffiltext{6}{NRC Fellow.}
\altaffiltext{7}{Miller Fellow.}


\begin{abstract}
We report on the design, first observing season, and analysis of data
from a new prototype millimeter-wave interferometer, MINT\@.
MINT consists of four 145\,GHz SIS mixers operating in double-sideband 
mode in a compact heterogeneous configuration.  The signal band is
subdivided by a monolithic channelizer, after which the
correlations between 
antennas are performed digitally. The typical receiver sensitivity 
in a 2\,GHz band is 1.4\,mK$\sqrt{\mbox{s}}$.  
The primary beams are 0.45\degr\ and 0.30\degr\ FWHM, with fringe
spacing as small as 0.1\degr.
MINT observed the cosmic microwave background (CMB) from Cerro Toco,
in the Chilean Altiplano.  The site quality at 145\,GHz is good, with
median nighttime atmospheric temperature of 9\,K at zenith (exclusive
of the CMB).  Repeated observations of Mars, Jupiter, and a
telescope-mounted calibration
source establish the phase and magnitude stability of the system.
MINT is the first interferometer dedicated to CMB studies to operate
above 50\,GHz.  The same type of system can be 
used to probe the Sunyaev-Zel'dovich effect in galaxy clusters
near the SZ null at 217\,GHz.  We give the essential features of MINT
and present an analysis of sideband-separated, digitally
sampled data recorded by the array. 
Based on 215 hours of data taken in late 2001, we set an upper
limit on the CMB anisotropy in a band of width $\Delta \ell=700$
around $\ell=1540$ of
$\delta T < 105\ \mu$K (95\% conf.). Increased sensitivity can be
achieved with more integration time, greater
bandwidth, and more elements.

\end{abstract}


\section{INTRODUCTION}
Measurements of the CMB at small angular scales ($\ell>1000$)
complement those from
{\sl WMAP} \citep{bennett03} and previous experiments including
TOCO  \citep{miller99}, 
BOOMERanG \citep{mauskopf/etal:2000,ruhl/etal:2003},
MAXIMA \citep{lee/etal:2001},
DASI \citep{halverson/etal:2002},
VSA \citep{grainge/etal:2003}, and
ARCHEOPS \citep{benoit/etal:2003},
allowing yet more precise constraints on the standard cosmological
model \citep{spergel03}.  Additionally, such measurements will enable us to
understand the transition from the linear to the non-linear growth
regimes for the formation of cosmic structure. One promising
measurement technique is interferometry because of the ability 
to control systematic errors. Existing CMB interferometers, 
CAT~\citep{robson93}, CBI~\citep{padin01}, DASI~\citep{leitch02}, 
IAC~\citep{dicker00}, VSA~\citep{watson03} are based 
on HEMT amplifiers~\citep{posp92,posp97,posp95} 
and operate at 30\,GHz 
where the contamination from point sources can be large;
MMIC technology at 90\,GHz will enable a straightforward
switch to higher frequencies.  A compact interferometer using SIS
(superconductor-insulator-superconductor) mixers at $f>100$\,GHz has
the potential 
to measure the CMB anisotropy and SZ effects where the point
source contamination is at a minimum~\citep{toffolatti} and where
the SZ~effect \citep{sunyaev70,sunyaev72} has a characteristic
spectral null. However, the technical complications are considerable,
as might be surmised by the existence of only a few such
interferometers (e.g., SMA, OVRO, BIMA, IRAM, NRO/NMA).

We have built and observed with a four element, heterogeneous,
compact, fully-digital, SIS-based D-band ($\sim 145$\,GHz) interferometer
designed as a prototype for a larger, more ambitious instrument.
Novel elements of the interferometer include the channelizer, a custom
microwave integrated circuit; digital correlators for CMB
measurements; and mixed antenna sizes.
We describe key features of the interferometer, 
the observations in Chile, and the data analysis.

\section{THE MILLIMETER INTERFEROMETER}

\begin{figure*}
\plotone{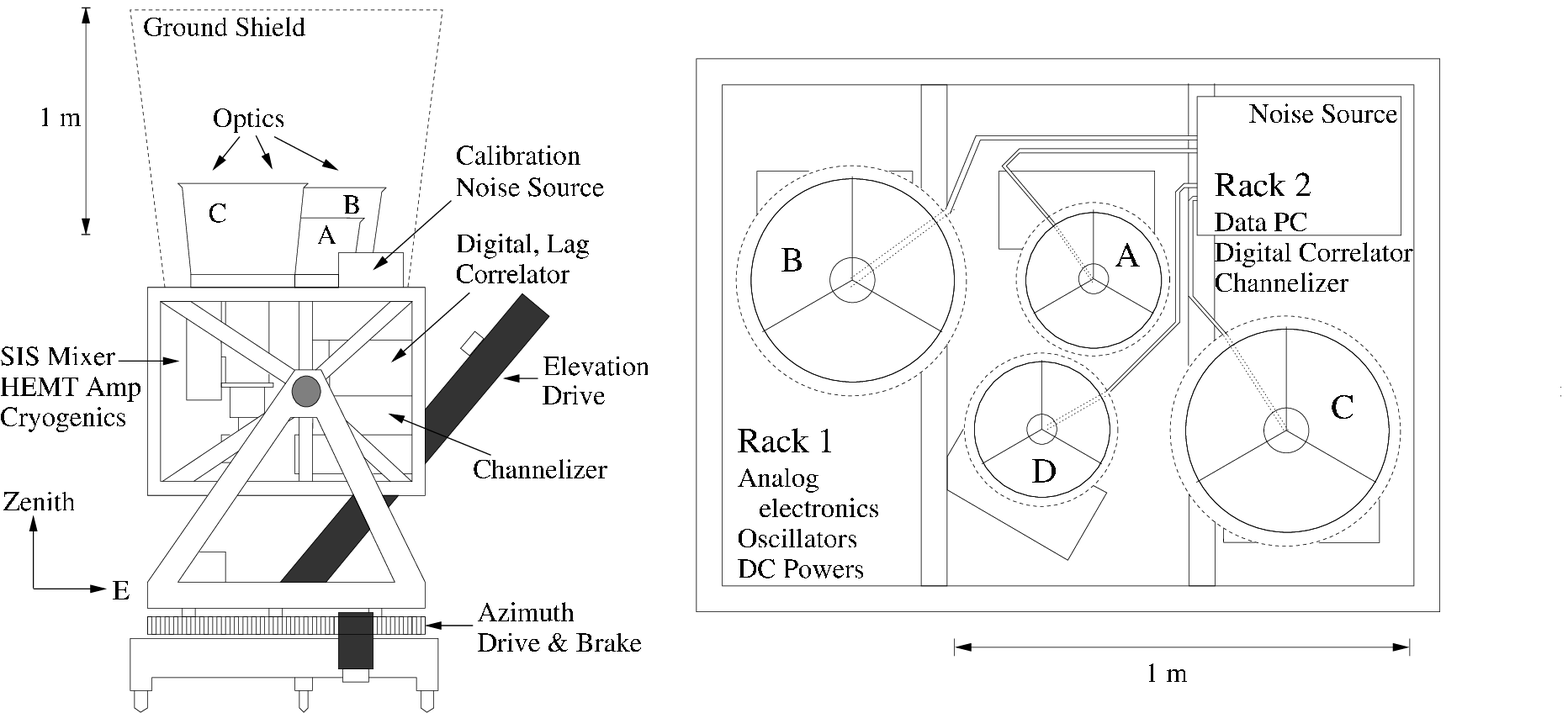}
\caption{\small MINT interferometer. 
An elevation over azimuth 
mount positions the cage that holds four Cassegrain antennas.
A ground shield  mounted on top of the cage moves with the receivers.
External to the cage, and not shown, are a thermal regulation 
system that servos the coolant for the RF electronics and 
correlators to $\pm1\,$K, a position control computer, the helium 
compressors, and a 50\,kW Katolight diesel generator to provide
electrical power. The entire experiment fits inside one 20-foot
shipping crate.
\label{fig:inst}}
\end{figure*}

The Millimeter INTerferometer (MINT) is a compact instrument, approximately
two meters on a side (without the ground shield) and cubical.  All four
receivers are mounted to a 
single rigid aluminum plate with fixed relative alignment.  The plate
sits atop a movable aluminum cage, as shown in
Figure~\ref{fig:inst}.  The channelizer, 
digital correlators, data computer, and a calibration noise source are
also mounted on the cage.  By moving the entire signal path
in unison, we avoid the potential phase changes that result
from cable flexure and other moving parts.

\ifnum\manuscript=1
\clearpage
\fi

\ifnum\manuscript=1
\clearpage
\fi

The telescope cage moves on an altitude-azimuth mount.
A linear actuator drives the telescope in elevation over the range
60\degr\ to 100\degr, controllable to the accuracy of the 17-bit
encoder ($\sim 0.005\degr$).  The azimuth drive has a range of 270\degr,
with an inaccessible zone centered to the north.  The
azimuth is also instrumented with a 17-bit encoder and can be
positioned to $\sim0.015\degr$ accuracy.  A brake on the azimuth
bearing allows us to turn off the electrically noisy azimuth drive
during fixed-azimuth observing.  A ground shield made of aluminum hex
cell in the
shape of an inverted frustum extends 1.2\,meters above the receiver
mounting plate.  This shield prevents the receivers from illuminating
the ground, either directly or through any number of reflections.

\subsection{Optics and Receiver Arrangement}

MINT uses antennas of two different sizes on its four receivers.
A heterogeneous array is unusual in radio interferometry.  It
introduces complications such as complex effective beams and an
``optical'' phase that are undesirable in most astronomy applications.
However, the faint CMB anisotropy is a weak signal at the limit of
detectability.  
We maximize the sensitive area in the aperture plane of the four-element
prototype array, within structural constraints, by combining antennas
of two different sizes.
Mounting the four antennas in a fixed configuration on a single plate
also improves sensitivity by ensuring that the baseline separations in
the ($u,v$) plane stay fixed as the sky rotates.  Of course, deep
observations of a single ($u,v$) visibility come at the expense of
mapping.

An on-axis Cassegrain mirror system is mounted above each receiver.
The inner two receivers have 30\,cm diameter primary mirrors and
6.5\,cm diameter hyperbolic secondaries.  The equivalent focal length
is 53.7\,cm, giving a focal ratio of $f/1.8$.  The outer
two receivers have Cassegrain optics of the same
shape but larger by a factor of 1.5 in each linear dimension.  Rays
from infinity focus near the entrance to a cold cylindrical feed
horn\footnote{Custom Microwave, Longmont, CO} inside the dewar.
Identical horns with 17\degr\ FWHM beams feed all four
antennas.  The feeds are characterized in \citet{Miller2002}.   
The mirrors were machined from aluminum on a computer-controlled
lathe.  A rigid tripod of thin (1.6\,mm) legs made of G-10 composite
supports each secondary mirror (see Figure~\ref{fig:inst}).  A conical
aluminum radiation shield 
with an opening half angle of 5\degr\ surrounds each primary mirror to
a height of approximately one diameter.  The shields are 
rolled outward at the top with a radius of $\sim 5\lambda$ to limit
diffraction and reduce crosstalk between receivers.

\begin{figure*}
\epsscale{1.1}
\plottwo{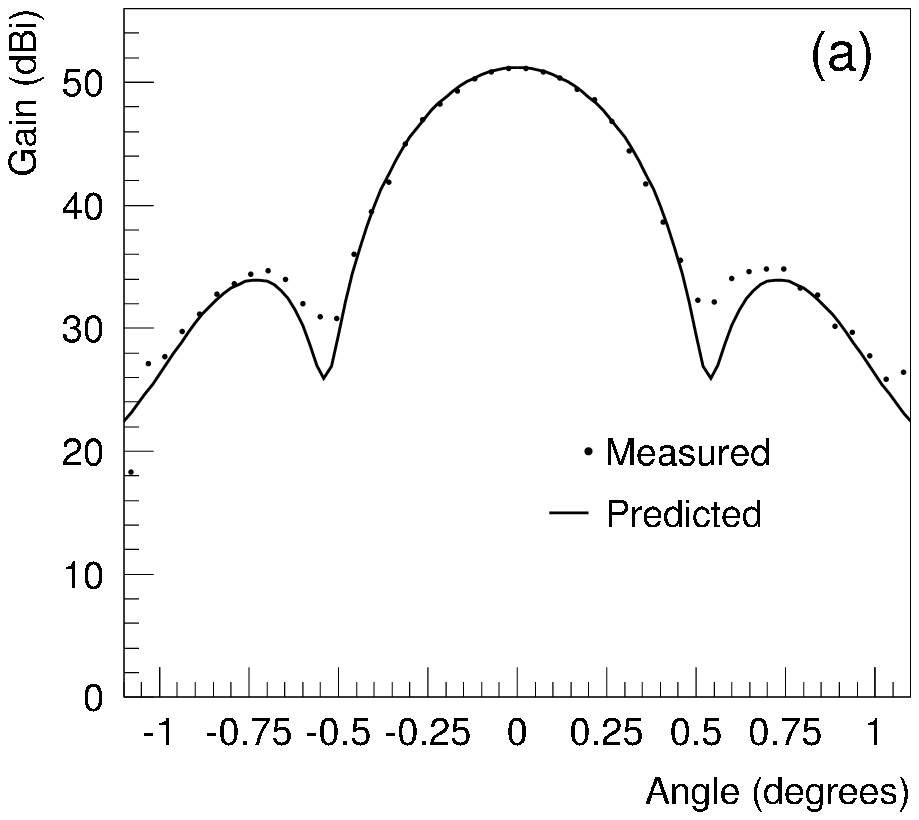}{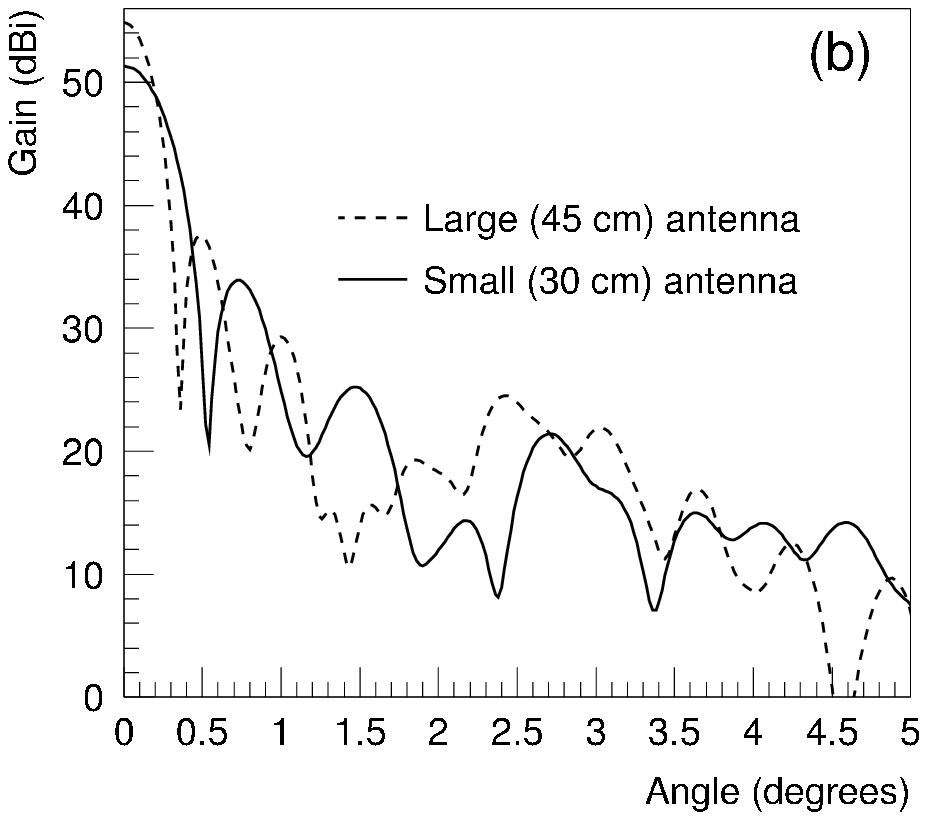}
\caption{\small 
(\emph{a}) Measured beam map of one of the small antennas. 
The dots are the measurement; the solid line is the DADRA 
computed beam, accounting for the defocus effect of measuring the beam at
finite distance (36 meters).  The data are normalized to the
computed forward gain.  Two-dimensional beam maps (not shown) confirm
that the beams of both antennas are circularly symmetric.
(\emph{b}) Far-field beams, as computed by DADRA, for the small
(solid) and large (dashed) antennas.}
\label{fig:beams}
\end{figure*}

The beam patterns of the two optical systems have been both measured
and computed.  Beam maps were measured using a chopped source (a
147\,GHz Gunn diode with a D-band corrugated feed) and a diode power
detector.   The system allowed measurements over a
dynamic range of 35\,dB\@.  
Beams were also computed using the DADRA 
program \citep{barnes02}.\footnote{YRS Associates:  Y. Rahmat-Samii,
W. Imbriale, \& 
V. Galindo-Israel 1995, rahmat@ee.ucla.edu} 
From a spherical wave expansion of the field from a 
corrugated feed, {DADRA} computes the currents on the 
subreflector. Then, from the subreflector, it computes the
currents on the primary. The net pattern is the sum of the fields
from the feed, subreflector, and primary. We found it necessary to
modify DADRA to include reflections from the primary back onto
the secondary, because the MINT secondary mirror has a
larger diameter than the hole in the primary.  The small primary hole
is desirable, as it puts power onto the sky rather than backward into the
dewar, but it must be included in the computation.  
Extensive measurements on the MINT and similar Cassegrain antennas 
(e.g., Figure~\ref{fig:beams}) confirm the computed beam shapes.

\ifnum\manuscript=1
\clearpage
\fi

\ifnum\manuscript=1
\clearpage
\fi

\subsection{Receivers}

\ifnum\manuscript=1
\clearpage
\fi

\begin{figure*}
\epsscale{0.9}
\plotone{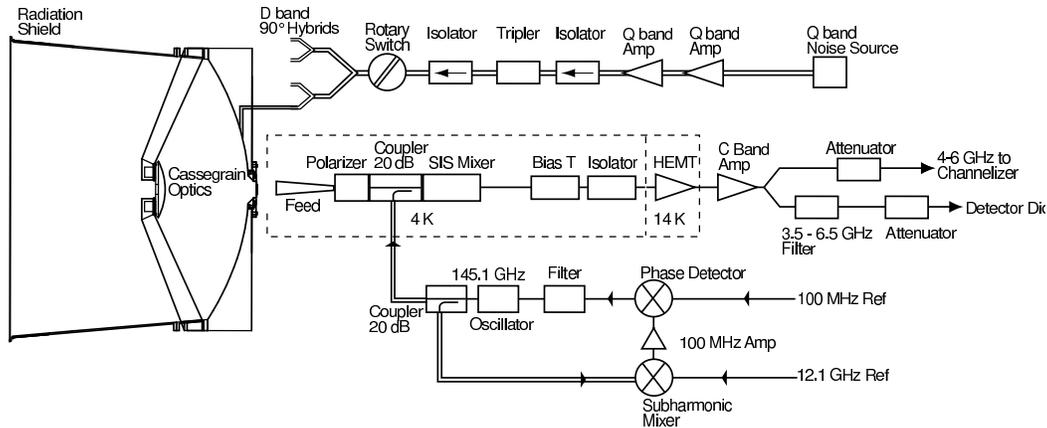}
\caption{\small Schematic of the MINT RF chain, calibrator, and phase
locking system.  The noise calibrator is the top section; the sky
signal flows from left to right in the center section; and the 145.1\,GHz
oscillator is produced and phase-locked in the bottom section.  The
dashed line surrounds cryogenically cooled components.} 
\label{fig:rec}
\end{figure*}

\ifnum\manuscript=1
\clearpage
\fi

The receiver is shown schematically in Figure~\ref{fig:rec}. It
uses the classic double-sideband superheterodyne configuration 
\cite[e.g.]{kraus86}. Other than the external optics, 
the four receivers are essentially the same.
Celestial radiation is focused by the Cassegrain optics onto a single
corrugated feed.  The feed couples to a polarizer that 
accepts only right circular polarization. The output of the 
polarizer supports linear TE$_{11}$ mode in a circular waveguide
(diameter = 0.173\,cm). A transformer \citep{padman77} converts it to 
TE$_{10}$ in WR-5.8.  The 1\,mW local oscillator\footnote{Zax Millimeter 
Wave Corporation, San Dimas, CA} (LO) at 145.1 GHz is coupled 
to the signal waveguide at $-20$\,dB through
a NRAO branch-line coupler \citep{kerr93}. On the SIS junction, the 
LO (300\,nW) and incident field ($\approx 0.2\,$nW)
are multiplied together.

For computing the optical transmission into the feed horn, the
ambient-temperature, 0.056\,cm poly\-pro\-py\-lene vacuum window; the
25\,K, 0.052\,cm thick z-cut quartz IR blocker; and the 4\,K backshort
are treated as a single multi-layer system. The spacing between the
elements and the thickness of the window and quartz were selected to
maximize the transmission near 145\,GHz.

Each SIS is cooled with a CTI-1020 CP refrigerator\footnote{CTI, Helix
Technology Corp., Mansfield, MA} modified with a third stage
following the design of \citet{plambeck93}. 
The modifications were made by Cryostar.\footnote{Cryostar Associates,
Yuma, AZ}
The resulting refrigerators are efficient; each has a cooling power of
0.05\,W at 4\,K when driven by a 2.2\,kW  CTI~8200 compressor.
The base temperature is approximately 3.7\,K,
but the temperature of the unloaded base varies by 400\,mK peak-to-peak
at 25\,Hz, the operating frequency of the fridge.
Reducing the variation with a high heat capacity, high conductivity
combination of stainless steel and copper
and judicious choice of thermal breaks, we were able to servo the 
cold head temperature to $\pm2$\,mK\@. This level of thermal stability
was essential. Even with such temperature
control, thermally driven dimensional changes in the cold head led to
a small modulation of the reflection of the LO power from the vacuum
window. The modulation could be monitored through $\sim0.5\%$
variations in the SIS bias current.  

The SIS was tuned with a backshort coupled through vacuum-tight gears 
to a computer-controlled motor outside of the dewar. 
The SIS is biased with a cryogenic NRAO bias-T, which
outputs the RF signal. Following
a 3.5--6.5 GHz cryogenic isolator,\footnote{Passive Microwave
Technology, Inc., Camarillo, CA and P \& H Laboratories, Simi Valley,
CA} a 4.1\,cm
piece of stainless steel waveguide leads to an NRAO 3.5--6.5 GHz,
5--8\,K noise temperature, 33\,dB C-Band HEMT-based amplifier at 14\,K\@. 
The output of the C-band amp feeds through stainless steel coaxial
cable to the room-temperature electronics
where it is amplified by another 52\,dB\@. Half the signal is coupled to 
a total power monitor and the other half goes to the channelizer.
With a 10\,K thermal load as input (similar to the load observed in the
field), the approximate level of the total power output is 1\,mW\@.

Each receiver is tuned while coupled to a cryogenic, thermally
controllable blackbody load that matches the feed but is electrically
and thermally isolated from it.  A LabVIEW computer program changes
the load temperature, SIS bias voltage, and backshort position 
to find the optimal bias. In addition to the stable SIS
temperature, the ambient electronics were 
stabilized to $\pm0.1$\,K\@.   Repeated tests of the system over
multiple cool-downs showed that
the optimal tuning position was stable. The same tuning procedure
worked without changes for all four receivers 
(A, B, C, D).  Table~\ref{tab:sis}
gives the noise temperature of the SIS system in the four bands.
After initial tuning, the SIS bias was not retuned during the
observing season.

To provide a continuous relative calibration, we built a 
138--142 GHz source that injects a broadband noise signal
toward the subreflector of each telescope (Figure~\ref{fig:rec}). 
A motor-controlled rotary mechanical switch ensured the ``off'' state
of the source. 

The LOs for all four receivers are phase locked to a 100\,MHz 
master clock, which also drives the digitizers and correlators,
the LOs in the channelizer and a 12.1\,GHz reference signal for
the LO locking. In the ambient-temperature electronics associated with
each  receiver, the 145.1\,GHz LO is phased locked to the 12.1\,MHz
tone using a circuit based on one given to us by the SMA
project \citep{hunter02} at the Harvard Smithsonian
Center for Astrophysics.  Each circuit is based on a type II
phase-locked loop, servoing on the twelfth
subharmonic of the 12.1\,GHz--145.1\,GHz mixing
product.\footnote{Pacific Millimeter Products, Golden, CO}  The LO
frequency is controlled by a varactor, an approach that requires far
less current than the more usual Gunn bias does.

To summarize, the outputs of each receiver are a phase-locked
4--6\,GHz IF signal and a total power monitor.  The receivers were
robust.  They operated for a total of 30 receiver-months without failure.
The receivers are described in more detail by \citet{randy_thesis}.

\ifnum\manuscript=1
\clearpage
\fi

\ifnum\manuscript=1
\begin{deluxetable}{lccccc}
\else
\begin{deluxetable*}{lccccc}
\fi
\tablecaption{Noise Temperature of Four Receivers Using NRAO SIS
mixers\tablenotemark{a}. 
\label{tab:sis}}
\tablewidth{0pt}
\tablehead{
\colhead{Band Name} &
\colhead{IF Band (GHz)} &
\colhead{$\rm T_A$ (K)}&
\colhead{$\rm T_B$ (K)}&
\colhead{$\rm T_C$ (K)}&
\colhead{$\rm T_D$ (K)}
}
\startdata
Blue   & 5.5--6.0 & 40 & 33 & 68 & 29 \\
Green  & 5.0--5.5 & 28 & 22 & 55 & 31 \\
Yellow & 4.5--5.0 & 24 & 19 & 33 & 27 \\
Red    & 4.0--4.5 & 26 & 19 & 32 & 24 \\

\tablenotetext{a}{Measured in the lab through the channelizer with a
regulated blackbody load.  The IF band is nominal band of the SIS output;
the bandwidth for each is close to 500\,MHz.}
\enddata

\ifnum\manuscript=1
\end{deluxetable}
\else
\end{deluxetable*}
\fi

\ifnum\manuscript=1
\clearpage
\fi

\subsection{The Channelizers}

The channelizers subdivide the 4--6\,GHz IF (intermediate frequency)
signal from each of the 
receivers into four 500\,MHz bands, then downconvert each band to
baseband.  The channelizers are custom integrated microwave
circuits designed using Agilent EEsof Advanced Design System
software.\footnote{Agilent Technologies, Palo
Alto, CA}  Each circuit consists of a 
specialized layout of etched copper on 0.076\,cm Duroid 6200 substrate
and drop-in components. Its only connectors (apart from power
supplies) are on the input and output.  The channelizers are housed in
a custom aluminum enclosure that provides isolation between adjacent
channels and from the outside
environment. The cavities between the traces and enclosure 
are coated with adhesive-backed, rubberized microwave absorber to 
suppress cavity modes. The channelizers for all receivers
are mounted as a single, compact (6\,cm by 30\,cm by 30\,cm),
temperature-controlled unit. 

The splitting is accomplished by a four-way power splitter
composed of planar cascaded Wilkinson power dividers followed by
planar bandpass filters. The Wilkinson splitters use a wide-band,
three-segmented design \citep{Li:1984}.  Each bandpass
filter is a six-pole Chebyshev type made from seven coupled
microstrip sections \citep{matthaei:1980}. 
The realized bandwidth varies from band to band, between 400 and
600\,MHz.  There is no significant variation among the nominally
identical bands in different channelizers.

Each filter is followed by a commercial surface-mount
mixer,\footnote{Mini-circuits, Brooklyn, NY} which downconverts the
signal to baseband (0--500\,MHz), and a surface-mount amplifier.  Two
local oscillator signals---one at 4.5\,GHz and one at 5.5\,GHz---are
sufficient to mix the four separate
bands to baseband. Thus half the bands are from the upper side-bands
of the LOs and half are from the lower. The overall gain of the channelizer is
$\sim$ 12\,dB across all four channels, with output power in
the range +1 to +7\,dBm.  We label the bands out of the channelizer
according to their IF frequency: red, yellow,
green, and blue. The red band is 4.0 to 4.5\,GHz in the IF, and thus
nominally contains celestial information from both sidebands of the
SIS: 149.1 to 149.6\,GHz and 140.6 to 141.1\,GHz.
\citet{tranthesis02} gives further information about the channelizers.

\subsection{The Digital Correlators}
\label{sec.correlators}

Each MINT digital correlator processes four 500\,MHz-wide analog input
signals, one from each receiver; all the red bands go to one
correlator, for example.  Incoming signals are digitized at 1 sample
per nanosecond with 2-bit resolution. The correlator produces a 16-lag
cross-correlation function, which is read into a data computer twice
per second. An offline Fourier transform of the 
correlation function produces a cross-power spectrum with 62.5\,MHz
resolution. A total of four correlators are required to process all
sub-bands. The basic design of the correlator was given to us by David
Hawkins \citep{hawkins}.

The electronics that compute the 16-lag cross-correlation
functions between four analog signals are designed onto a single 38\,cm
by 38\,cm, 8-layer, surface-mount circuit board.   The four
incoming signals are first processed by identical RF sections
that each contain a variable attenuator, an amplifier, a low pass
filter, and a voltage bias.  The variable attenuators 
(M/A-COM model AT20-0107) have smallest steps of 0.5\,dB\@.  They are
adjusted by data acquisition software to maintain the rms signal
level into the digitizer near 0.28\,V\@.  This
level optimizes signal-to-noise given our correlator algorithm
and the digitizer quantization at $\pm 0.25$\,V \citep{cooper}.
Resetting the attenuators hourly was found in field tests to be
sufficient.  The low-pass filters (Mini-circuits SCLF-550) have a 
$-3$\,dB point at 605\,MHz and block unwanted noise
(particularly leakage of the channelizer LOs), which otherwise would
be aliased into the signal.  The voltage bias of $-0.5$\,V is required
by the digitizers.

The RF pre-processor is followed by a commercial 1\,GSPS, 6-bit
digitizer (SPT7610).\footnote{Signal Processing Technologies, Colorado
Springs, CO}  The two most significant bits are passed
to an ECL demultiplexer, which slows the signal rate to 62.5 MHz on a
32-bit bus.  After conversion to LVCMOS, the 32-bit data
streams from each receiver are fed into a single FPGA (Field
Programmable Gate Array, XCV1000),\footnote{Xilinx Inc., San
Jose, CA} which computes the 
correlation functions and communicates with a digital I/O card in the
data PC\@.  The FPGA also keeps a running count of the digitized values
for each receiver---the
``digitizer histograms.''  These histograms, which are read out twice
per second, proved very useful as monitors of the RF power
reaching the digital correlator (see \S\ref{sec.abs-cal}).

The algorithm for the correlator is built
around a multiplier-adder kernel.  The multiplier is implemented as a
4-bit lookup table, whose values are given by the 4-level, ``deleted
middle products'' correlator algorithm.  This table has a
quantization efficiency of $\epsilon_q= 0.88$ relative to an
ideal analog correlator \citep{cooper}.

The rack-mounted correlators for all receivers fit into a box $41$\,cm
by $41\,$cm by $20\,$cm.  Together the four consume 800\,W and are cooled
with a thermally regulated fluid.  The correlators 
worked as designed throughout the 2001 campaign.  More details are
given by \citet{tranthesis02}.

\section{OBSERVATIONS}

\subsection{Cerro Toco Site}
Observations were made from an altitude of 5200\,m on Cerro
Toco\footnote{ The Cerro Toco site of the Universidad Cat\'olica 
de Chile was made available through the generosity of Prof. Hernan
Quintana, Dept. of Astronomy and Astrophysics.}
(longitude $67^\circ47'09''$ West, latitude $22^\circ57'29''$ South),
a mountain overlooking the high Chilean Altiplano near the ALMA and
CBI sites.  This was the same site used for the MAT/TOCO experiment
\citep{Torbet99,miller99}.  The mountain is accessible from San Pedro
de Atacama, Chile, by four-wheel drive vehicle and can be reached from
Princeton, New Jersey in 24 hours.

The atmosphere at the Cerro Toco site was often excellent.  During the
38 calendar days of the campaign (2001 November 22 through December
29), twenty percent of potential observation nights were lost to
overcast weather or snow.  On the
remaining usable nights, the effective atmospheric temperature at
D-band was below 9\,K half the available time.  The first and third
quartiles were 7\,K and 13\,K and the minimum observed temperature was
5\,K\@.  The typical opacity in D-band was therefore only
a few percent.  These temperatures exclude the contribution of the CMB
and cover only the period from 23:30 to 10:00~UT each night (local
time is UT$-3$).  Data totaling 215 hours are used in the
analysis presented here.

Two radio communication systems and automated observing software
generally permitted remote operation of MINT from an apartment in the
nearby town of San Pedro, 35\,km away \citep{Miller2002}.  One-way data
transmission from the site (33 kbps data rate) allowed observers to
monitor the data in real time.  A separate, two-way, low-bandwidth
system employed packet modems to relay commands to the telescope
control computer.

\subsection{Absolute Calibration}
\label{sec.abs-cal}

The digital correlator measures the quantity $\rho(\tau)$, the
unitless correlation coefficient of two voltage signals as a function
of lag $\tau$:
\begin{displaymath}
\rho(\tau)\equiv
\frac{\langle V_a(t)V_b(t+\tau)\rangle }{\sqrt{\langle V_a^2(t)\rangle
\langle V_b^2(t)\rangle }},
\end{displaymath}
where brackets represent averages over time $t$.
The correlator output is converted from raw bits to $\rho$
following \S8.3 of \citet{tms}.  The next conversion, from $\rho$ to
correlated power or temperatures, requires knowledge of
the overall system temperature of each receiver.

The most robust power calibration comes from the total power diodes,
which measure power at the IF in a 3\,GHz band (wider than but
inclusive of the signal band).  Small nonlinearities in the diodes
are characterized in the
lab.  The linearized voltage is calibrated in the field to the input
temperature, using a combination of hot/cold load tests and sky dips.
The hot load is a large piece of Eccosorb at ambient temperature
($T\approx 270$\,K); the cold load is the sky at zenith.  By measuring
the sky power at a range of elevations, we separate the sky
temperature, the receiver temperature, and the receiver responsivity
(volts per Kelvin).  The receiver temperatures and responsivities
found in this way were consistent in three separate measurements taken
at Cerro Toco over a three-week period.  Assuming that responsivities
remain constant, each diode serves as an independent monitor
of the sky temperature.  In practice, all four diodes found the same
sky temperature to within 0.8\,K rms, confirming that no other
important contribution to the measured power varied significantly
during the campaign.

The sky (atmosphere plus CMB) serves as a variable-temperature load,
given the Eccosorb and sky-dip calibrations.  
The digitizer histograms give the power in
each 0.5\,GHz-wide sub-band.  Together, the power and sky temperature
permit an estimation of the receiver temperature and responsivity of
each sub-band.  Although the responsivity is found to change in one or
a few discrete jumps during the campaign for some bands, the noise
temperatures do not change significantly.  
Thus the total system temperature in each channel can be
determined at any time simply by adding this constant receiver noise
temperature to the varying sky temperature.  The jumps, of approximately
10--20\% in magnitude, occur only between nights and are attributed to
problems in signal cables or their connectors.

This calibration method assumes that several key receiver
properties stay constant: the responsivity of the total power diodes;
the receiver temperature at the diodes; and the receiver temperature
in each 0.5\,GHz-wide sub-band.  The redundant measurements of the sky
temperature confirm these assumptions \citep{randy_thesis}.

Observations of Mars and Jupiter are used to establish the forward
gain (including optical efficiency) of the receivers.  The gain is
required to convert between brightness temperature and the
fluctuations in the CMB physical temperature.  Planetary observations
are described in Section~\ref{sec.planets}.

\subsection{Relative Calibration and Phase Stability}
\label{sec.noise_source}

The phase and amplitude stability of the interferometer are monitored
using a switchable noise source.  The noise is produced by
a +20dB ENR Q-band source\footnote{Noise Com,  Parsippany, NJ} 
followed by a series of amplifiers providing a gain of +70\,dB\@.  This
signal enters an isolator, a frequency tripler, and another isolator
(Figure~\ref{fig:rec}).  A D-band rotary waveguide  
switch (controlled by a stepper motor) allows the noise
to be turned on or off.  The signal is then split four ways by a
$180^\circ$ magic-T hybrid coupler and a pair of $90^\circ$ short
slot hybrid couplers.  The resulting
four signals are highly correlated, except that each is phase-delayed
by $90^\circ$ relative to the next.  All components up to and
including the power splitters are housed in a temperature-regulated
box.  Each signal travels through one
meter of coin-silver D-band waveguide, where it is coupled into the
telescope optics by an open-ended waveguide illuminating a tiny ``pickoff''
mirror adjacent to the dewar window.  Some power reflects up to the
secondary and back into the dewar.
The coupling efficiency is not measured but is found to be stable.
The noise source is sharply cut off above 145\,GHz by the
bandpass of the Q-band amplifier, with the consequence that the noise
contains D-band power only in the MINT lower sideband ($139-141$\,GHz).
This property is helpful in verifying the sideband separation procedure
(\S\ref{sec.sbsep}).

During normal CMB observation, the noise source is switched on for 6--8
seconds out of every 450.  A longer calibration of 30 seconds is
performed every hour.   The noise source produces power equivalent to
5--10\,K in each receiver, so the correlated signal is observed with
high signal-to-noise in a single 2-second integration period.

The magnitude of the noise signal recorded by the interferometer was
stable throughout the campaign.  On most nights, the fractional range
of amplitudes was 3\% (rms) or less in all bands and all 
baselines.  Over the entire campaign, the rms range was 6\%.  Strong
correlations in the amplitudes among all 
bands and all receivers suggest that the main source of amplitude
variation was the noise source itself.  For example, some variation is
due to the waveguide switch failing to find the same resting spot on each
repetition.

The phase stability of the instrument---a critical requirement for any
interferometer---was also measured using the noise source and was found
to be sufficient.  Considered over the
campaign, the measured phase variation over time and across frequency bins
would cause average signal loss of 6\% in most baselines (corresponding to
an rms phase variation of 20\degr).
The D-C and A-B baselines performed better than the others, with
average loss of less than 3\% and $\phi_{rms}=12\degr$.  We attribute
this difference to the power-splitting scheme in the noise source (A
and B signals come from the same 90\degr\ hybrid), which permits a
relative phase between the A/B signals and the C/D ones.  This
observation suggests that the calibration source itself is responsible
for the main part of the observed phase variation over time.
Therefore, the interferometer 
is more phase-stable than could be measured with this source (a
comment that also applies to amplitude stability), and
the 12\degr\ rms phase variation is an upper limit on the
stability of the interferometer itself.

The instrument and sky noise is observed to integrate down with rms
signal values scaling 
as $t^{-1/2}$ for averaging times $t$ of up to 2000 seconds in a
single night.  Longer integrations were too few to make significant
comparisons with the expected scaling law.

\section{OBSERVATIONS OF THE SUN, MARS AND JUPITER}
\label{sec.planets}

Repeated observations of the Sun were used to determine the telescope
alignment throughout the Chile campaign.  The receivers
were temporarily covered with standard corrugated cardboard 2\,mm thick
to block the Sun's optical and infrared radiation.  The pointing was
derived from separate maps of each receiver's total power, not from
interferometry.  Common-mode pointing errors can be corrected off line
and were determined to approximately 0.025\degr\ accuracy in elevation
and in azimuth.  Following an initial alignment of the receivers to each
other (which also involved solar observations), they were found to be
relatively aligned to within $\sim0.02\degr$.  This angle corresponds
to approximately 1/6 of the fringe spacing in the longest baseline.
We noted no significant shifts
during the campaign.  At this level of accuracy, the flatness and
placement of the cardboard shields appear to limit the measurements.

During the Chile campaign, Mars was visible nightly in the late 
afternoon and early evening.  When weather and maintenance
permitted, MINT observed Mars by repeatedly pointing the telescope
ahead of Mars' position and allowing the planet to drift through the
beam for 1--2 minutes.  Approximately twenty hours of data were taken
on or near Mars.  Averaged over the season, the planet's angular
diameter was $7\arcsec$.  Using the 171\,GHz temperature \citep{goldin} of
198\,K, we expect Mars to produce antenna temperature in the
small-small, mixed-size, and large-large baselines of only 1.9\,mK,
2.9\,mK, and 4.3\,mK (the three baselines have forward gains of
+51.3\,dB, +53.1\,dB, and +54.80\,dB, according to the lossless optical
model, and solid angles of 93\usr, 62\usr, and 42\usr).  Jupiter was
observed only at the end of the campaign.

The Mars data are used to measure the phase versus frequency profile
of the instrument.  Since Mars appears as an unresolved point source
to MINT, its 
visibility phase should be a simple function of its position relative
to the phase center of a given baseline.  In particular, the phase
should be nearly independent of the base-band frequency (effects of
wavelength variation are small due to the small fractional bandwidth
at D band).  With the phase corrected for the position of Mars, a
fiducial phase $\phi(\nu)$ is determined by averaging the complex
visibility over many Mars observations.  All data, including planetary
calibrations and CMB data, are adjusted by subtracting the fiducial
phase.  After making this adjustment, it is possible to average
complex visibility across all frequencies.


The visibilities measured on Mars match the important features of a
point source.  The visibility magnitude is approximately Gaussian (as
a function of pointing angle away from the planet) with a width as
expected from the primary beam widths.   The visibility phase is also
consistent with the expected $2\pi \mbox{\boldmath $x \cdot u$}$ fringes.  The
phase of the setting Mars is stable over the campaign at the level of
5\degr--10\degr\ rms, depending on the baseline.  Even in the worst
baseline, the computed signal loss due to the phase variation is only
3\%.  The phase is somewhat less stable during
observations of Mars rising (10\degr--20\degr\ rms, or 2\%--10\% loss),
but these data are taken during daylight
hours and are therefore less indicative of the instrument stability
during its nighttime CMB observations.

Mars' visibility magnitude, though stable during the campaign, does not
match the expected magnitude.  For all baselines involving
receiver~A, the magnitude is only 33\% of the expected value; for
other baselines, the observation is 75\% of the expected value.
Extensive tests to explain the deficit were performed after MINT
returned to Princeton in 2002.  The tests involved direct observation
of Jupiter, a high signal-to-noise source, and confirmed the signal
loss.  
By switching equipment between the receivers, channelizers, and 
correlators, the deficit was localized to the poor performance of the
circular-to-linear polarizers at the base of the feeds.
The degradation is clearly evident only while observing a known 
celestial source interferometrically (which was not possible prior to
deployment).  This loss of coherent signal is corrected at the end of
the analysis pipeline by scaling up all average and rms values.

Jupiter was too low in the northern sky during late 2001 to
be visible to MINT in its normal configuration.  After the CMB
campaign ended, the ground shield was removed and replaced by a flat
mirror that covered all receivers.  This mirror redirected the beams
of each receiver by 76\degr\ in altitude, allowing
observation of Jupiter.  The planet's larger angular diameter ($47\arcsec$)
makes the expected signal much larger than the Mars signal and easily
discerned in a single 2-second data frame.    We assume a brightness
temperature of 172\,K for Jupiter, given by {\sl WMAP} measurements at 90\,GHz
\citep{page03}.  The Jupiter observations (taken on consecutive
mornings 2002 January 2--4) confirm all the main results of the lower
signal-to-noise Mars data, including phase and amplitude stability,
beam sizes, and the reduction in signal (particularly in
receiver A) caused by the polarizers.  

\section{OBSERVING THE ANISOTROPY}

\ifnum\manuscript=1
\begin{deluxetable}{crr}
\else
\begin{deluxetable*}{crr}
\fi
\tablecaption{Median right ascension\tablenotemark{a} \ and galactic coordinates for CMB
fields with at least 9 hours observing time\tablenotemark{b}.
\label{tab:galactic}}
\tablewidth{0pt}
\tablehead{
\colhead{RA (h,m,s)} &
\colhead{$b$ (deg)} &
\colhead{$l$ (deg)}
}
\startdata
\ra{1}{3}{42}  & $-84.47$ & 161.59 \\
\ra{2}{3}{5}   & $-72.68$ & 202.49 \\
\ra{3}{2}{40}  & $-59.54$ & 212.27 \\
\ra{4}{2}{34}  & $-46.34$ & 218.46 \\
\ra{5}{2}{26}  & $-33.12$ & 223.82 \\
\ra{6}{2}{16}  & $-20.24$ & 229.17 \\
\ra{7}{2}{8}   &  $-7.68$ & 234.97 \\
\ra{8}{2}{6}   &    4.37  & 241.77 \\
\ra{9}{2}{1}   &   15.52  & 249.94 \\
\ra{10}{1}{47} &   25.41  & 260.21 \\
\ra{11}{1}{39} &   33.31  & 273.18 \\
\ra{12}{1}{38} &   38.34  & 289.07 \\

\tablenotetext{a}{The declination of each strip is $23\degr 2\arcmin
14\arcsec$\ south.}
\tablenotetext{b}{The nominal fields are spaced evenly in right
ascension.  The coordinates given here are medians of the 
data surviving all cuts.}
\enddata

\ifnum\manuscript=1
\end{deluxetable}
\else
\end{deluxetable*}
\fi

\subsection{Sky Scan Strategy}
An experiment's sensitivity to CMB anisotropy depends on how the
observing time is divided among fields on the sky.  The statistically
optimal division requires observing each field until a
signal-to-noise ratio of one is achieved, then moving on to the next.
For MINT, this meant maximizing the observing time on each field.  We
used a quasi-tracking strategy which combines the stability advantages
of the sky drifting over a stationary telescope with increased
integration time beyond a normal drift-scan.  The telescope was
repointed eight times per hour to observe the same strip of sky
(1/8 hour wide in right ascension) repeatedly.  Pointings were made
at a fixed azimuth angle and a range of near-zenith elevations.
Although fixed points in the sky curve across a range of azimuth, this
effect was small for the limited range of elevations used.

Table~\ref{tab:galactic} gives the location of the center of each
strip used in the current analysis, both in ecliptic and in galactic
coordinates.  Although the fields near RA=$7^\mathrm{h}$ and
$8^\mathrm{h}$ are within 10\degr\ of the galactic plane, they are
located far from the galactic center and outside other major
components.  Additionally, the spectrum of diffuse
backgrounds such as interstellar dust falls approximately as
$\ell^{-1.5}$ \citep{gautier:1992}.
The galactic contribution in the MINT fields at multipoles
$\ell\gtrsim1000$ is negligible.

The effect of scanning continuous strips is to acquire a series of
complex visibilities with signal highly correlated between neighbors.
Simulations show that sky strips  $w=1.875\degr$ long observed with a
Gaussian primary beam of width $\sigma_b=0.19\degr$ have the same
statistical power for measuring sky variance as $n_{spot}=4.5$ separate
spots.  The general relation for all three MINT beam sizes
is found to be 
$$n_{spot}\approx1+\frac{w}{2.8\sigma_b}.$$

\subsection{Data Selection}
The data used for CMB analysis are selected based on time of day, sky
temperature, and nominal receiver operation.  The noise source
calibrations (\S\ref{sec.noise_source}) show that the system
phase response is stable from 23:35~UT until
10:05~UT\@.  These times correspond approximately to half an hour after
sunset and twenty minutes after sunrise.  Outside these times, the phase
variations are likely due to thermal changes in the noise calibration
waveguide (which is insulated but not thermally regulated).  Although
we believe that the instrument is more stable than the calibrator,
there is no way to be certain, and data are rejected outside this
window of 10.5 hours per night.

In addition, the data are cut when the sky brightness in D-band
(including CMB) exceeds an effective temperature of 35\,K\@.   At higher
temperatures, the linear relationship between sky temperature and
measured rms visibilities breaks down.  The sky brightness cut removes
3\% of the otherwise valid data, which due to their high noise
temperature would have only 1\% of the statistical weight.

Instrument criteria include tests for locked PLLs, functional 4\,K
refrigeration, and properly synchronized correlator readout.   The
total data set after all cuts is 215 hours long, representing 54\% of
the available nighttime hours.  The weather and the instrument are
responsible for approximately equal shares of the lost time.

\section{ANALYSIS}
\subsection{Expected CMB Signal for a Heterogeneous Interferometer}

MINT has both 30\,cm- and 45\,cm-diameter antennas, a configuration
which produces high 
collecting area for the medium and longest
baselines, where the CMB signal is small, while 
permitting the shortest baseline to be only 160\,$\lambda$.
Put another way, the heterogeneous array fills a larger fraction
of the antenna plane than would an array of small antennas alone.

Given a CMB power spectrum, the expected variance of the
interferometer visibility is derived by \citet{lasenby} and
\citet{white99} for the case of identical receivers.
We consider here the calculation for a heterogeneous array, which
differs in that the effective beam and its Fourier transform are
complex.

We work with a modified ``temperature visibility'' quantity, which
corresponds to a brightness temperature averaged over the beam solid
angle, rather than the usual flux density.  This definition amounts to
a re-scaling of the usual visibility by a constant factor
$\lambda^2/(2k\Omega_B)$, where $\Omega_B$ is the beam solid angle.
We make this non-standard choice even though MINT's power levels are
ultimately calibrated to the flux density received from point sources
(Mars and Jupiter).  We have found it convenient to work in sky
brightness temperature, a quantity appropriate to a beam-filling
source.  In particular, this choice permits quick and transparent
comparison with non-interferometric  CMB experiments and their
sensitivities.  We use the label $V_T$ to designate the modified
visibility definition; all formulas for $V_T$ can be multiplied by
$2k\Omega_B/\lambda^2$ to yield the conventional visibility.

Ignoring the frequency dependence of the CMB
brightness across MINT's narrow band and making the flat-sky
approximation for small fields of view, the complex temperature
visibility $V_T$ at location $\bfu$ in the $u-v$ plane is
\begin{equation}
\label{eq.vis_temp}
V_T(\bfu)\equiv\int\frac{ \dif\x}{\Omega_B}A(\x)\Delta T_B(\x)e^{+2\pi
i\scriptstyle \bfu\bfcdot \x}, 
\end{equation}
where \x\ is the two-dimensional position in the plane of the sky
(conjugate to \bfu), $A$ is the primary power beam of the antennas
(normalized to one at its peak),  
$\Omega_B\equiv\int d\x\, A(\x)$ is the beam solid angle,  
and $\Delta T_B$ is the CMB brightness temperature fluctuation ($\Delta
T_B=\etarj \Delta T$).  The Rayleigh-Jeans
factor \etarj\ converts small physical temperature
fluctuations $\Delta T$ to brightness temperatures:
\begin{displaymath}
\etarj\equiv\frac{\partial T_B}{\partial T}=
\frac{\lambda^2}{2k}\ \frac{\partial B}{\partial T}
=\frac{x^2 e^x}{(e^x-1)^2},
\end{displaymath}
where $x\equiv h\nu/(kT)$.
At  145\,GHz and for $T=2.73$\,K, $\etarj=0.59$.  
Note that the definition of temperature visibility
(equation~\ref{eq.vis_temp}) leads to a magnitude of
$|V_T|=T_B\,\Omega/\Omega_B$ for an unresolved source of brightness
temperature $T_B$ and solid angle $\Omega\ll\Omega_B$.  This fact
connects the CMB observations to the absolute calibration off of Mars.

Apart from constant
factors, the visibility is the Fourier transform of the product of $A$
and $\Delta T_B$\@.  By the convolution theorem, it is also the
convolution of their transforms: 
\begin{displaymath}
V_T(\bfu)=\frac{\etarj}{\Omega_B}(\tilde A \otimes
\widetilde{\Delta T} )(\bfu).
\end{displaymath}
Using this expression for temperature visibility, its expected covariance
matrix has diagonal elements of
\begin{eqnarray}
C^T &\equiv&   \langle V_T^\ast(\bfu)V_T(\bfu)\rangle \\
&=&\left ( \frac{\etarj}{\Omega_B} \right)^2
\int\,\dif\bfv\, |\tilde A(\bfu-\bfv) |^2\,T_{cmb}^2\,S(\bfv) \label{eq.ct_def}
\end{eqnarray}
using the unitless power spectrum, $S(\bfv)$, defined by
\begin{displaymath}
\langle\tilde T^\ast(\bfv) \tilde T(\bfu)\rangle =
T_{cmb}^2\,S(\bfv)\delta(\bfv-\bfu).
\end{displaymath}
Note that the Fourier power spectrum $S$ depends only on the magnitude
of $\bfv$ and approximately
equals the spherical harmonic spectrum 
$C_\ell$ for large $\ell$ (if the spectrum is smooth enough):
$S(v)\approx C_{\ell=2\pi v}$. 

Non-diagonal elements of the theory covariance matrix are of two types.
Two observations made by a single baseline but separated by
angle ${\bfDelta}$ on the sky are accommodated with a factor
$\exp(2\pi i{\bfDelta}\bfcdot\bfv)$ in the integrand of
Equation~\ref{eq.ct_def}.  Observations made by distinct baselines are
also correlated.  In the extreme case of parallel baselines (such as
DA-BC and AC-DB in MINT), the correlation is complete.  The MINT
baselines of equal length but different orientations (such as DB and
DC) are weakly correlated.  The ratio of the magnitude of
the covariance for DB-DC data  to that of the DB-DB data is 18\% for
data from a single field ($\Delta=0$), falling off quickly with
increasing field separation $\Delta$.  The correlations between the
short or the long baselines (DA or CB) and any other are negligibly
small.

Analysis of the data is more easily done using the real and imaginary
visibilities separately, leading to the final covariance elements
\begin{equation}
\label{eq.crr}
C_{RR}^T= \frac{(\etarj T_{cmb})^2}{2\Omega_B^2}
\int\,\dif \bfv\, |\tilde A(\bfu-\bfv) |^2 S(\bfv)
\cos(2\pi{\bfDelta}\bfcdot\bfv)
\end{equation}
and
\begin{equation}
\label{eq.cri}
C_{RI}^T= \frac{(\etarj T_{cmb})^2}{2\Omega_B^2}
\int\,\dif \bfv\, |\tilde A(\bfu-\bfv) |^2 S(\bfv)
\sin(2\pi{\bfDelta}\bfcdot\bfv).
\end{equation}
The other combinations obey $C_{II}^T=C_{RR}^T$ and
$C_{IR}^T=-C_{RI}^T$.
The new factor of 2 accounts for the fact that half of the
sky variance appears in the real component of $V$ and half in the
imaginary.  The two formulas are even and odd with respect to exchange
of the two indexes (or equivalently, under reflection of 
\bfDelta).  This result can be generalized to give the covariance
between distinct baselines.  

In words, the covariance for $\bfDelta=0$ is the sky power
spectrum $S(v)$ averaged in two dimensions with a weighting function
$|\tilde A(\bfu-\bfv) |^2$.  Recall that $A(\x)$ is a single receiver's
power beam, so $\tilde A(\bfu-\bfv)$ is the autocorrelation of the
voltage pattern in the antenna plane, offset from the origin by a
distance $\bfu$ corresponding to the baseline separation in wavelengths.

The complication of mixed antennas can be handled by extending the
formulation of the antenna power reception pattern~\citep{tms}.
Starting instead with the voltage pattern $g_j({\x})$ of
antenna $j$ for radiation from angle \x,
the effective power beam $A_{e}$ for baseline $j$-$k$ is 
\begin{displaymath}
A_{e}({\x}) = g_j({\x}) g^\ast_k({\x}).
\end{displaymath}
With this change, all formulas and statements above about $A$ and
$\tilde{A}$ hold, 
except that $\tilde{A}(\bfu-\bfv)$ becomes the \emph{cross}-correlation
of the two antenna's voltage patterns in the aperture plane.
The resulting beam $A_e$ is, in general, complex.  The overall phase is
unimportant; it merely adds to any other phase offsets in the 
system, which are measured by observing planetary point sources. 
Phase variation across the beam, however, adds an imaginary
part to the beam $A$ and its Fourier transform $\tilde{A}$.\@
In MINT's mixed-antenna baselines, the phase of the effective beam is
nearly constant well past 
the half-power point of the main lobe, so the imaginary part of
$\tilde{A}$ is of much smaller magnitude than the real part.  Even so,
the imaginary part is included in $|\tilde{A}|^2$ when numerically
integrating equations \ref{eq.crr} and \ref{eq.cri}, increasing the
theory covariance by a few percent.

\subsection{Approximate Analytic Expression for the Expected Signal}

The theory covariance integrals can be approximated by taking the 
effective beams to be Gaussian and the power spectrum to be
scale-invariant.  The result is useful in illustrating how the CMB
signal depends on instrument parameters.   We assume a Gaussian beam
with a beam solid angle $\Omega_G=2\pi\sigma^2$.  
The ratio of the beam solid angles for the  idealized Gaussian to the
actual beam is akin to a main beam efficiency; call it
$\epsilon_g\equiv 2\pi\sigma^2/\Omega_B$.  The transform $\tilde A$ of the
normalized beam is then
\begin{displaymath}
\tilde A(\bfu)= 2\pi\sigma^2\,e^{-2(\pi\sigma u)^2}.
\end{displaymath}
 Let
\begin{displaymath}
S(v)=\frac{1}{2\pi v^2}\left( \frac{\delta
T^2}{T_{cmb}^2}\right ),
\end{displaymath}
so that $\delta T^2$ is a flat-bandpower temperature variance.
Performing the angular part of the integral over
$\dif \bfv$ (Equation \ref{eq.ct_def}) yields
\begin{displaymath}
C^T_{RR}=\frac{(\etarj\,\epsilon_g\,\delta T)^2}{2}
\int\frac{\dif v}{v}e^{-(2\pi\sigma)^2(u^2+v^2)}
I_0(8\pi^2\sigma^2 uv),
\end{displaymath}
where $I_0$ is the modified Bessel function of the first kind
\citep{dieguez,hobson:2002}. 
This can be simplified with the asymptotic expansion $I_0(x)\approx
e^x/\sqrt{2\pi x}$ for large $x$.  Defining $U\equiv 2\pi \sigma u$ and
$V\equiv 2\pi\sigma v$,
\begin{displaymath}
C^T_{RR}=\frac{(\etarj\,\epsilon_g\,\delta T)^2}{2 U^2}
\int \dif V\,\frac{1}{\sqrt{4\pi}}\left(\frac{U}{V}\right)^{3/2}
e^{-(U-V)^2}.
\end{displaymath}
Note that $r_P=\lambda/(2\pi\sigma)$ is the characteristic radius of the
Gaussian power reception pattern in the antenna plane.  Therefore,
$U=\lambda u/r_P> 2$ for any real interferometer, as the
separation $\lambda u$ must be larger than the antenna diameter.
The integrand above is small except near $V\approx U$, so the
divergence as $V\rightarrow 0$ can be ignored (it corresponds to an
unphysical divergence in the CMB power spectrum at large angular
scales, where the flat-sky approximation breaks down as well).  For
$U\approx3.5$, the integral evaluates numerically to 0.54, approaching
0.500 in the limit of large $U$ (large separation).  Replacing the
Gaussian beam width $\sigma$ with the  beam full width at
half-maximum power $\theta_\mathit{FWHM}=\sqrt{8\ln2}\,\sigma$ (which
is more readily generalized to non-Gaussian beams),
the simplified estimate for the covariance matrix is
\begin{eqnarray}
C^T_{RR} &\approx& 0.27\, \etarj^2\,\epsilon_g^2\, \delta T^2\ U^{-2} \\
 &\approx& 0.038\ \etarj^2\,\epsilon_g^2\,\delta
 T^2/(u\theta_\mathit{FWHM})^2.\label{eq.gauss_estimate} 
\end{eqnarray}
Equation~\ref{eq.gauss_estimate} can be used to estimate the
magnitude of the CMB signal from any single interferometer baseline,
given basic parameters such as baseline length and the beam width and
efficiency.  We expect that the ``Gaussian efficiency'' $\epsilon_g$
can generally be replaced with the main beam efficiency with little
loss of accuracy.

\subsection{Expected Signal-to-Noise on the Power Spectrum}
A numerical evaluation of equations~\ref{eq.crr} and \ref{eq.cri} gives a more
accurate figure for the expected MINT variance per $\mu\mbox{K}^2$ of
flat bandpower (Table~\ref{tab.results}).  The two-dimensional Fourier
transform of the beam, $\tilde A$, is simplified by noting that the
antennas and beams are symmetric about the optical axis.  This
symmetry reduces the problem to the one-dimensional Hankel transform
of the radial function.  The numerical result assumes the beam
shapes computed by physical optics and confirmed by direct measurement
(Figure~\ref{fig:beams}).  

The noise covariance $C^N$  follows straightforwardly from the system
temperature and the  Dicke equation~\citep{dicke}.  The covariance of
the real or imaginary output of a complex correlating interferometer is
\begin{displaymath}
C^N\equiv \langle \Delta T_B^2\rangle =
\frac{T_{sys}^2}{2\epsilon_q^2 \Delta \nu \tau},
\end{displaymath}
where $T_{sys}$ is the system noise temperature,
 $\epsilon_q$ is the efficiency factor of the digital quantization scheme,
 $\Delta\nu$ is the IF bandwidth, and $\tau$ is the effective
observing time (see for example \citet{tms} eq.~6.42).  Note that
for MINT, $\tau$ is only half the elapsed time, because the
phase-switching scheme effectively time-shares the correlator between
measuring the real and imaginary parts of the correlation.
Using a typical
$T_{sys}=45$\,K (including atmosphere and CMB), $\epsilon_q=0.88$,
$\Delta\nu=1.9$\,GHz, and 5 hours of observing time ($\tau=9000$\,s) on
a CMB field, the noise covariance is 77\,$\mu\rm{K}^2$.  

Using a set of $N\gg 1$ Gaussian deviates to estimate the
variance of the set gives a fractional uncertainty of $\sqrt{2/N}$.  The
number of samples $N$ is twice the number of fields observed, $n_f$,
because the complex correlation at each field gives two independent
samples.  The variance $C^T+C^N$ of the data can be estimated to
a fractional uncertainty of $1/\sqrt{n_f}$, so
\begin{displaymath}
\frac{\delta C^T}{C^T} \approx \frac{1}{\sqrt{n_f}}\left (
\frac{C^T+C^N}{C^T}\right).
\end{displaymath}
The two terms are the sample and noise variances.
The quantity of interest, $\delta T$, is an rms rather than a
variance; thus if the error is small, $\delta T$ has half the
fractional uncertainty of $C^T$.\ 
Consider the case of MINT observations, covering $n_f=50$ fields.   If sample
and noise variance contributed equally to the final error (which is
statistically optimal for a fixed total observing time), then the
fractional error expected on the CMB signal $\delta T$ would be
$1/\sqrt{50}\approx15\%$.  Achieving this level of uncertainty would
require 76 hours of observing per field, approximately 12 times the
time actually available in the 2001 campaign.

\subsection{Sideband Separation}
\label{sec.sbsep}

The SIS mixers in MINT are inherently double-sideband devices, mixing
sky signals in both ranges 139--141\,GHz and 149--151\,GHz down to a
single IF band at 4--6\,GHz.  By switching a 90\degr\ phase shift in
and out of the 145.1\,GHz local oscillator tone every half-second, we
separate the signals from the two sidebands.   The
separation is possible because an LO phase shift enters with opposite
signs in the IF signals from the two sidebands.  If $C$ and $C'$ are the
correlation functions for unshifted and shifted data, then $(C+iC')$
and $(C-iC')$ isolate the contributions from lower and upper
sidebands.  This feature is preserved through the channelizer, where
the signals are mixed down a second time.  \citet{tms} describe the
technique in \S6.1.  We have shown that their result generalizes  to
the cases of unequal power in the two sidebands and of lower sideband
conversion at the second mixing stage.

The magnitude and phase of the cosmic signal should be nearly the same
in both MINT sidebands.\footnote{The lower sideband power exceeds the
upper by 7\% for CMB fluctuations, owing to the frequency dependence
in the conversion from physical to brightness temperatures.}
Therefore, it is not strictly necessary for CMB  observation to
distinguish the signals in the two bands.   We could maintain a
constant LO phase, effectively measuring only one component of the
complex sky visibility.  Instead, MINT determines the full complex
visibility with half the observing time per component.  The MINT
strategy was chosen to characterize the instrument completely and to
guard against possible drifts in the instrumental phase.   Correcting
such phase drifts---if measured by the calibration source
(\S~\ref{sec.noise_source})---requires the complex visibility.

The sideband separation technique is proved in practice by
observations of the calibration noise source.  The noise source
produces power only in the lower sideband of the MINT receivers.  The
measured cross-power 
spectra of the calibrator shows no power in the upper sideband
except for small leakage near the band edges, a result of the limited
number of time lags available in the correlation function.

\ifnum\manuscript=1
\clearpage
\fi

\ifnum\manuscript=1
\begin{deluxetable}{lccccccc}
\else
\begin{deluxetable*}{lccccccc}
\fi
\tabletypesize{\scriptsize}
\tablecaption{MINT Baselines, Bandpower Conversion Factors, and
Results of 2001 Campaign.\tablenotemark{a} 
\label{tab.results}}
\tablewidth{\textwidth}
\tablehead{
\colhead{Baseline } &
\colhead{$u$} &
\colhead{\parbox[b]{5em}{\raggedright 
Beam solid angle $\Omega_B$ ($\mu$sr)}} &
\colhead{\parbox[b]{5em}{\raggedright 
Beam FWHM $\theta_\mathit{FWHM}$~(deg)}} &
\colhead{\parbox[b]{8em}{\raggedright Conversion factor,
Gaussian ($10^{-3}\  \mu\mathrm{K}^2/ \mu\mathrm{K}^2$)}} &
\colhead{\parbox[b]{8em}{\raggedright Conversion factor,
Numerical ($10^{-3}\  \mu\mathrm{K}^2/ \mu\mathrm{K}^2$)}} &
\colhead{\parbox[b]{6em}{\raggedright Effective $\ell$ for flat $\delta T$}} &
\colhead{ \parbox[b]{8em}{\raggedright Upper limit on $\delta T$ ($\mu$K, 95\% C.L.)} }
}

\startdata
Short                  & 169 & 93 & 0.445 & 4.3 & 4.3 & 980 &232\\
Mixed\tablenotemark{b} & 260 & 62 & 0.348 & 2.5 & 2.4 &1540 &105\\
Long                   & 492 & 42 & 0.297 & 1.1 & 1.0 &3030 &323\\

\tablenotetext{a}{
The two conversion factors give the ratio of signal variance (in
brightness temperature) to the CMB bandpower $C^T/\delta T^2$.  For example, a
bandpower of $1000\,\mu\mathrm{K}^2$ would produce a signal of 
$4.3\,\mu\mathrm{K}^2$ in the short baseline.
Gaussian and Numerical columns refer respectively
to the simplified estimate (Equation~\ref{eq.gauss_estimate} in the
text) and to the full numerical integration of Equation~\ref{eq.crr}
used for the analysis.
The conversions include several factors.
The Rayleigh-Jeans factor squared is $\sim 1/3$.
Signals separated into real and imaginary parts each contain only
half of the anisotropy power.  Use of a realistic beam instead of a ``top
hat'' introduces a further factor of two in signal.  
Finally, the largest contribution, a factor of 15--60, comes from 
incomplete sampling of the $u$-$v$ plane.  The effective $\ell$ is the
mean value probed if the power spectrum is flat.
The last column gives the upper limits found by MINT on the CMB
anisotropy.
}
\tablenotetext{b}{The conversion factors are per baseline.  MINT has
four baselines of the intermediate length $u=260$.}
\enddata

\ifnum\manuscript=1
\end{deluxetable}
\else
\end{deluxetable*}
\fi

\ifnum\manuscript=1
\clearpage
\fi

\section{ANISOTROPY RESULTS}
\subsection{Data Reduction}

Analysis of the interferometer data is performed in Fourier
(temporal frequency) space.  First, the correlation function $\rho(\tau)$ is
converted to brightness temperature (as described in
\S\ref{sec.abs-cal} and \S\ref{sec.planets}).  The conversion
incorporates the results of sky dips and the absolute calibration
determined from Mars observations.  Using 
the sideband separation procedure, data with and without the 90\degr\
phase shift in the LO are combined to form a complex correlation
function.  The discrete Fourier transform of the 16-lag correlation
gives the cross-power spectrum of one pair of antennas.
Interpretation of the cross-power is complicated by the fact that both
the lowest and highest frequency bins combine signal from upper and
lower sidebands.  This complication is not a problem, because MINT
does not currently make use of the spectral information in analysis of
the CMB data.

The ``fiducial phase'' $\phi(\nu_j)$ found from many Mars observations
(\S\ref{sec.planets}) is subtracted from each data point.  The
subtraction accounts for the constant phase variation across frequency
of components such as imperfectly matched waveguides and amplifiers.
The visibility magnitude is also divided by a template (based on
planetary observations) to correct for gain variation across the
signal band.  With 16 lags and four separate 500-MHz sub-bands, the
template for each baseline has 64 complex elements, which are
determined once for the whole campaign by averaging the Mars results.
Phase- and magnitude-corrected data are now directly comparable, and
are therefore averaged across all 64 frequencies.

After all cuts, there are 390,000 two-second data records.
Each record is a complex visibility (in brightness temperature) for each
of the six baselines.  The data show non-zero averages over the
campaign of approximately 5--20\,$\mu$K\@.  These offsets do not vary
significantly with time or with the observing elevation, so the six
complex offsets are subtracted from the data records.  Cross-talk
between signal channels is the most likely cause of the small offsets.

CMB data are combined into bins of width 0.1\degr\ in right ascension
(the bins are smaller in actual sky angle by a declination factor of
$\cos\delta=0.921$).  All data are taken at fixed declination, so
sky binning is required only in one dimension.  The bin size is chosen
to be the largest bin with nearly complete correlation of
signal visibilities across its width.  The visibility of a CMB
signal varies across the bin with the usual $2\pi\bfu\bfcdot\x$ phase,
but this position dependence is removed by ``phase-centering'' the
visibility as if it had been observed at the center of the bin.
Given a width of 0.1\degr, the signal lost due to averaging is only
1.5\% in the longest baseline and 1.2\% and 0.8\% in the others, a
loss which is accounted for by scaling the final results.  The signal
reduction would be closer to 10\% for most baselines if the
phase-centering correction were not made (the exception being
the two baselines with \bfu\ oriented north-south).

Combining the data within a bin entails finding the weighted average
and the uncertainty on the mean, as given in~\citet{lasenby}.  
Each data point has a noise variance given by the 15-minute average of
the visibility variance, scaled by the instantaneous system
temperature to account for variation in the sky temperature on faster
time scales.  The rms estimated from this procedure matches the rms of
the phase-centered data to better than 2\% for all baselines, when
taken over the entire campaign.  For the final likelihood analysis,
the estimated ``instantaneous rms'' is scaled to eliminate the small
mismatch.  

Data from the identical baselines DC and BA could be averaged
together at this point, but the large loss in Receiver~A makes the BA
data subject 
to unknown systematics and far less sensitive than the DC data;  the
AC data is likewise less sensitive than the BD data.  We
therefore discard the results from the BA and AC baselines.

\subsection{Results}
The time-averaged visibilities at each bin are our best estimate of
the complex sky visibility on a 0.1\degr\ R.A. spacing.  A maximum
likelihood analysis (e.g., \citet{bondjaffeknox,hobson:2002}) is used
to find the flat   bandpower in three distinct bands, corresponding to
the MINT baseline lengths.  The signal covariance matrix, given by
equations \ref{eq.crr} and \ref{eq.cri}, is computed numerically for
the appropriate power spectrum ($S(v)\propto v^{-2}$) for all six
baselines, and for all possible separations
$\Delta$ from zero to 3\degr\ in 0.1\degr\ steps.  For this
computation, the beam transform $\tilde A(u)$ comes from the physical
optics code.  The DB-DC correlation of 18\% is included in the
covariance matrix, but the small
covariances between  baselines of unequal length
($<1\%$ in all cases) are ignored.  Because the scan strategy
concentrated on 2\degr\ strips spaced every 15\degr, the signal
covariance is block diagonal with approximately 18 separate blocks
(of the 24 possible blocks, six correspond to areas of the sky high
during the afternoon, which were not observed).  The noise
covariance matrix is diagonal. 
These simplifications make the total covariance matrix easy to invert.

Table~\ref{tab.results} gives the 95\% confidence level upper limits
determined by MINT in the 2001 campaign.  We compute the likelihood
$\cal L$ as a function of flat bandpowers $\delta T$,  given a uniform
prior distribution.  The three 
baseline lengths are analyzed separately, as the correlations between
them are in this case small enough to ignore.  The separate likelihood
functions of the mixed baselines are added, as their data are
uncorrelated in $u$-$v$ space.  We define the upper
limit so that the integral of $\cal L$ up to the limit is 95\% of
the integral over the whole domain.  The mixed baselines
($\ell\sim1540$) provide the strongest limit, 105\,$\mu$K, on the CMB
anisotropy.

\section{CONCLUSION}
The Millimeter Interferometer has operated successfully as a prototype for
a high-frequency CMB instrument.  MINT combines heterogeneous
antennas, double-sideband SIS mixers, monolithic channelizers,
and digital correlators.  All of these elements may be features
in future CMB interferometers.  To make a competitive detection of the
CMB anisotropy, such  
an instrument would require wider bandwidth, longer observing time, and
more receivers (having more distinct baselines would also improve the
sampling in $\ell$-space).  
Usable bandwidth in the D-band is ultimately
limited by atmospheric emission lines \citep{matsushita}, but
sideband-separating mixers---under development at NRAO
\citep{sb_sep_sis}---offer an instant factor of two improvement, given
equal system noise temperatures.  
Consider an improved MINT, with 8\,GHz bandwidth (4\,GHz upper
and 4\,GHz lower sideband), four times the actual 2001 observing time,
and fully working polarizers.  Assuming 50 separate fields of view and the
power spectrum of the 2002 WMAP best fit $\Lambda$CDM model
\citep{spergel03}, the MINT window functions produce CMB bandpowers of
1450 and 590 $\mu\mathrm{K}^2$ at effective $\ell=900$ and $\ell=1390$.
In this case, the improved MINT would have the
raw sensitivity to see anisotropy at the level of $7\sigma$ and
$6\sigma$ in the short and medium baselines.  We conclude that the
techniques described here constitute a promising, 
if challenging, method for observing CMB anisotropies at or above
100\,GHz.

\acknowledgments
We gratefully acknowledge design and construction help from Dave
Wilkinson, Norm Jarosik, Ted Griffith, Laszlo Vargas, Bill Dix, Glenn
Atkinson, Stan Chidzik, 
Elvis Dieguez, Jamie Hinderks, Mark Tygert, Michael Levi, William
Magrabe, Ariel Lazier, Long Tran, 
Mark Morales, Paul Oreto, and Michael Nolta.  
MINT was made possible by SIS mixers designed by Tony Kerr, 
phase-locked loops modified from a design of Bob Wilson and Robert Kimberk, 
cold heads designed by Dick Plambeck, 
and a digital correlator modified from a design of David Hawkins.
These colleagues freely offered design advice and guidance.
Agilent provided a complimentary educational license for EEsof design
software.  Mark Devlin loaned us his telemetry system.
Angela Glenn and Kathleen Warren assisted in countless ways,
which included purchasing equipment, producing
this document, and getting MINT and the authors to Chile.
Operating in Chile was possible through the efforts of Angel
Ot\'{a}rola and Roberto Rivera.  Use of the Cerro Toco site was made
possible by Hernan Quintana.  For
additional assistance in Chile, we thank 
Gaelen Marsden, Juan Burwell, the entire CBI team (particularly for
help when our truck broke down), Ocegtel SA, and the staff of La
Casa de Don Tom\'as.
We thank an anonymous referee for comments which helped us to improve
the paper.
This work was supported by an NSF NYI award,
a Robert H. Dicke Fellowship to JF,
a David and Lucile Packard Fellowship to LP, 
an NSF Graduate Fellowship to TM,
and NSF grants PHY96-00015 and PHY00-99493.

\ifnum\manuscript=1
\pagebreak
\fi

\end{document}